# The Political Economy of FDI flows into Developing Countries: Does the depth of International Trade Agreements Matter? †

Arslan Tariq RANA[*]
arslan@univ-orleans.fr

Mazen KEBEWAR[*]
mazen.kebewar@univ-orlenas.fr

**Abstract**: There is considerable debate whether the domestic political institutions (specifically, the country's level of democracy) of the host developing country toward foreign investors are effective in establishing the credibility of commitments are still underway, researchers have also analyzed the effect of international institutions such as (GATT-WTO) membership and Bilateral Investment treaties (BIT) in their role of establishing the credibility of commitment to attract foreign investments. In addition, most recent studies have examined the effect of International Trade Agreements (TAs) on FDI flows as they contain separate investment chapters and dispute settlement mechanism, thus providing confidence to investor regarding the security of their investments. We argue that there are qualitative differences among various types of trade agreements and full-fledged trade agreements (FTA-CU) provide credibility to foreign investors and democracy level in the host country conditions this effect whereas the partial scope agreements (PSA) are not sufficient in providing credibility of commitments and not moderated by democracy. This paper analyses the impact of heterogeneous TAs, and their interaction with domestic institutions, on FDI inflows. Statistical analyses for 122 developing countries from 1970 to 2005 support this argument. The method adopted relies on fixed effects estimator which is robust to control endogeneity on a large panel dataset. The strict erogeneity of results by using a method suggested by Baier and Bergstrand (2007) and no feedback effect found in sample. The results state that (1) More the FTA-CU concluded, larger the amount of FDI inflows are attracted into the developing countries and PSA are insignificant in determining the FDI inflow; (2) FTA-CU are complementary to democratic regime whereas the conditional effect of PSA with democracy on levels of FDI inflows is insignificant.

† Acknowledgments: A preliminary version of this paper was presented in European Association of Comparative Economic Studies; we would like to thank the participants for their comments and suggestions, which have significantly improved the paper. We would like also to thank Tim Bûthe for sharing the data with us.

* University of Orleans (France) – Faculty of Law, Economics and Management – Orleans Economic Laboratory (LEO), UMR CNRS 7322, Rue de Blois, B.P. 26739 – 45067 Orléans Cedex 2 – France.

# 1. Introduction

Foreign Direct Investment (FDI) inflows have recently been the most persistent source of capital inflows in developing countries. In 2010, these countries have attracted FDI inflows of $574 billion constituting the share of 46.29% of total inward investment (UNCTAD 2011). This extensive growth of FDI has, consequently, given rise to the competition among policy makers in developing countries that adopt higher investment incentives and make ex ante commitments to foreign investors about the continuity of policies. The instruments they use to make commitments are the membership in international organizations such as GATT-WTO. Other instruments are conclusion of Bilateral Investment Treaties (BIT) and Preferential Trade Agreements (PTAs). The conclusion of PTAs is considered a new phenomenon in establishing credibility of commitment in the goal to attract FDI flows (e.g. Büthe and Milner 2008). Apart from international institutions, the domestic politics also plays a crucial role and the regime type conditions the impact of PTAs in attracting FDI (Büthe and Milner 2010). PTAs are not homogenous. There exist numerous types of PTAs. WTO distinguished among their various types. The focus here is on the differing impact of various types of trade agreements on FDI flows and also to examine whether the democratic regime moderates the effect of different types on FDI inflows.

The governments are confronted with obsolescing bargain in the process of attracting foreign investments. In the obsolescing bargain, the investors (MNCs) are in a position to locate operations in one of the multiple possible locations around the world. To attract investments, governments offer variety of attractive investment incentives to the foreign firms in the form of reduced taxes, investment grants and wage subsidies (Li and Resnick 2003). But FDI is often characterized by a high degree of reversibility and sunk costs (Stasavage 2002b, and Jensen 2003). Thus, the investors are worried about the credibility of the attractive stances taken by the governments and the successful continuation of the policies after the investment is made. Examples of policy changes include changes to performance requirements, direct expropriation, nationalization, and confiscation of foreign assets (Henisz 2000).

The host developing countries can signal their credibility of commitments at domestic as well as international level. Numerous researchers have found positive impact of a democratic regime on FDI (e.g. Busse 2003). According to Jensen (2003), democratic political institutions constrain the policy flexibility of the executive thus signaling to foreign investors about policy stability after FDI location. But Li (2009) opposed and argued that democracy is not a remedy for eliminating risk of expropriation. There exists contradiction in literature about the impact of democratic regime on FDI. While debates on whether domestic political institutions (specifically, the country's level of democracy) of the host developing country toward foreign investors are effective in establishing the credibility of commitments are still underway, researchers have also analyzed the effect of international institutions such as (GATT-WTO) membership and Bilateral Investment treaties (BIT in their role of establishing the credibility of commitment to attract foreign investments. In addition, most recent studies have examined the effect of International Trade Agreements (PTAs) on FDI flows (Büthe and Milner 2008, 2010) as they contain separate investment chapters and dispute settlement mechanism, thus providing confidence to investor regarding the security of their investments. These studies analyzed this issue based on PTAs dummy variable without considering that not all PTAs provide the same level of investment protection and liberalization. A deeper analysis is needed to take into account the qualitative differences among various types of trade agreements.

This paper builds on the study of Büthe and Milner (2010), who examined the effect of interaction of regime type with PTAs, and analyzed the impact of heterogeneous PTAs, and their interaction with domestic institutions, on FDI inflows to examine the conditional effect. Full-fledged agreements such as Free Trade Agreements and Customs Unions (FTA-CU) serve as a better commitment device as they are associated with domestic political as well as international reputational cost, whereas Partial Scope Agreements (PSA) have limited credibility ascertaining no obvious cost in case of reneging. Each type consists of different arrangements and thus differs in its institutional characteristics. Statistical analysis for 122 developing countries from 1970 to 2005 is done. The method adopted relies on fixed effects estimator which is also robust to control endogeneity on a large panel dataset. The strict exogeneity of results is tested by using a method suggested by Baier and Bergstrand (2007).

The rest of the paper is structures as follows: Section (2) presents the review of literature, analyzing the effect of heterogeneous PTAs and discussing the contradiction in impact of regime type on FDI flows. The simultaneous effect of varying types of PTAs and level of democracy is also examined in this section. Section (3) includes explanation of data and results, whereas section (4) concludes.

## 2. Theoretical Background of interaction of institutions

### 2.1. Heterogeneity of International Institutions and FDI

At the international level, countries, competing for international production capital, have concluded the establishment of the institutions which bind them of their commitments signaling the partner countries, the persistence of the policies and protecting the rights of the foreign companies making (long term) investments in the host country. The benefit of international institutions lies primarily in the creation of disincentives for states to behave opportunistically by reneging on trade agreements (Goldstein and Martin 2000). Among them, the multilateral institution is GATT-WTO, which has provided the forum for investment provisions and its protection. Apart from the multilateral institutions, there exist regional and bilateral arrangements known as PTAs. Although, varying in the coverage of products with the commitment for tariff reduction, range of issues and entailing the measures for the implementation of those commitments, PTAs offer twofold advantages in the context of attracting FDI: PTAs include strategic instruments such as provisions for investment[1], reducing barriers to investment and transaction costs. They provide dynamic benefits such as increased bilateral investment, industrial location and help in growth (Schiff and Winters 1998) and secondly, at the same time, they serve as the insurance device and provide mechanisms for making credible commitments to foreign investors about the continuation of policies and treatment of their assets, thus avoiding the time-inconsistency trap, reassuring investors and increased investment (Fernandez and Portes 1998, Simmons 2000, and Büthe and Milner 2008). In other words, trade agreements institutionalize the investment process (from negotiations on investment provisions till the mechanisms to protect them after dislocation of firms). In addition to the investment and assurance benefits to investors, PTAs, in general offer lower trade barriers, increase the scope of products to be traded and indirectly decreased costs for investors.

In the light of the above discussion about the advantages of PTAs in attracting production capital, they nevertheless differ in scope and coverage as well as in provision mechanisms and legalization. Trade agreements vary widely in depth and form. WTO distinguishes the agreements according to their form such as partial scope agreements, free trade agreements, customs unions, to which common markets can be added. These agreements range from the simple exchange of trade preferences on limited range of products and investment measures to the harmonization of policies beyond tariffs such as dispute settlement procedures, legality, competition policy, infrastructure and environmental standards. GATT-WTO under the article XXIV permits the formation of FTA under the condition if the partner countries liberalize "substantially all trade" among them. Customs unions are a one step further to FTA entailing common external tariff for all the member countries as introduced by Balassa (1961), considering regionalism as gradual process. On the other hand, Partial scope agreements (in which there is little coverage and the level of legalism is very low) are permitted for developing countries under the "enabling clause" to exchange partial tariff preference between members.

---

[1] Although trade agreements are negotiated for trade related issues and are intended to increase trade flows, growing number of trade agreements contain investment issues.

The heterogeneity between the institutions entailing different depth and breadth are the drivers of diverse investment outcomes. The reason is due to the involvement of varying provisions regarding investment and dispute settlement, punishment mechanisms and their legal enforceability in trade agreements negotiations. States, while negotiating various agreements can include third party for arbitration (such as International Centre for Settlement of Investment Disputes) in case of dispute emergence, specifically related to investments[2]. Emphasizing on the differing effects of regional integration agreements on investments, Blomström and Kokko (1997) argue that perhaps the most serious challenge facing a study of the relation between regional integration and foreign direct investment is the multi-dimensional character of the issue. As argued, the diversity of trade agreements has differing effects, they not only reduce barriers to trade and investment, and they also provide increased benefits relating to the implementation of credibility of commitment.

Trade agreements are in variety of forms and the type of agreement at hand influences the type of dispute settlement system favored by member governments (Smith 2000). Highly legalized dispute settlement mechanism indicates a strong commitment to the protection of property rights of investors and, subsequently, should be an effective way of attracting investment flows (Büthe and Milner 2008). Also, the effectiveness of an integration agreement depends upon the punishment mechanisms and costs it implies for pursuing bad policy (such as reneging) (Schiff and Winters 1998). Institutions may have relatively strong central authorities and significant operating responsibilities or be little more than forums for consultation (Koremenos *et al.* 2001), the later implying a very low level of legalism permitting the concluding state to reject any proposed settlement in favor to accommodate domestic interest groups and caring about policy discretion to craft policies (Smith 2000). No doubt, the low level of legalism still provides investment and assurance benefits to some extent, there are two broader advantages inherent in deeper and broader arrangements. Firstly they generate larger economic gains by freeing 'substantially' all trade and thus investments and secondly by involvement of strong dispute settlement mechanisms and options, they provide greater predictable environment for investors who seek to raise their voice in the event of policy change (ex-post agreement) or expropriation of their assets in the host country after the location of their investments.

A full-fledged trade agreement[3] serves as the better commitment device than partial scope agreement[4]. A strong, liberalizing RIA may lead to a virtuous circle of increased credibility and thus increased investment and growth. On the contrary, a more closed agreement could lead in the opposite direction with reduced credibility and lower investment (Schiff and Winters 1998). For example Abbott (2000) establishes that though EU and NAFTA[5] adopted different institutional and juridical models, they preferred 'hard' law. The broader and deeper agreements such as FTA-CU (and not PSA which involve weaker obligations), ascertaining strongly binding legal commitments, offer advantages to both parties signing an agreement with the objective to increase the flow of investments. First, the hard legal commitments are used by the government of investors or MNC as assurance devices as the potential for opportunism and costs are high for the host countries (Abbot and Snidal 2000). Secondly, for the host countries[6], they serve as mechanism to prove their credibility of commitments about the treatment of foreign investors' assets when noncompliance is difficult to assess ex-ante. Focusing on the preferences of countries for using hard law in international economic arrangements, Abbott (2000) suggests that hard law reduces private risk premiums and intergovernmental transaction costs associated with trade and investment. Thus, reduced uncertainty and costs associated with broader and deeper forms of economic integration augments the flow of investments among the signing parties. Figure (1) depicts the evolution of Full-fledged trade agreements (FTA-CU) as well as partial scope agreements (PSA) of developing countries and annual flows of FDI into these countries.

---

[2] Currently, there are 157 signatory States to ICSID convention (http://icsid.worldbank.org).
[3] Although, there is debate that even recent Free Trade Agreements do not offer hundred percent free trade as states do often exclude sensitive sectors such as agriculture, still here we treat them as full-fledged agreements here as they are still better than partial scope arrangements in liberalizing trade and investment and legal enforcement of resolving disputes.

The design of agreement proposed prior to an arrangement also signals the extent of credibility. If the design of arrangement is flexible and weakly legalized where asymmetry of information is high, such as in (PSA), the plausibility of reneging on the agreement cannot be ignored. Goldstein and Martin (2000) argued that although flexibility or "imperfection" could lead to stability and success of agreements, but incentives also exist for states to evade commitments. Also, Abbott and Snidal (1998) argued that noncompliance typically results not from deliberate cheating but from ambiguity in agreements. There exists potential that as these loosely integrated (flexible) agreements can be reneged and thus, in turn, they do not provide a strong mechanism to provide confidence to investors. Figure 1 provides the evolution of FTA-CU, PSA and Annual FDI inflows.

These arguments regarding the heterogeneity of trade agreements and flows of foreign direct investments yield the testable hypothesis:

It can be seen from the figure that FTA-CU and FDI inflows are growing together whereas PSA are showing different trend although the figure shows their similarity with FDI inflows before the year of 1995.

H1: The greater number of FTA-CU a country concludes the greater and significant will be FDI inflows in contrast to the number of PSA concluded.

FDI, in the wake trade agreements, have domestic implications for governments. Therefore, it is important to examine how they react to the varying institutional setups of trade agreements in their ability to attract FDI flows.

---

[4] Kahler (2000) argues hard legalization is less significant in looser economic integration and in which the states do not pursue deep integration such as APEC (Asia-Pacific Economic Cooperation), therefore not providing assurance of sincere commitments. APEC is characterized by low levels of precision, delegation, and obligation and not a successful model of regional economic integration (Abbott 2000).
[5] According to WTO, NAFTA is recognized as FTA and EC/EU as CU.
[6] Developing countries in general, as their domestic institutions and political conditions are not stable as compared to developed economies.

**Figure (1)** FTA-CU, PSA and FDI Inflows

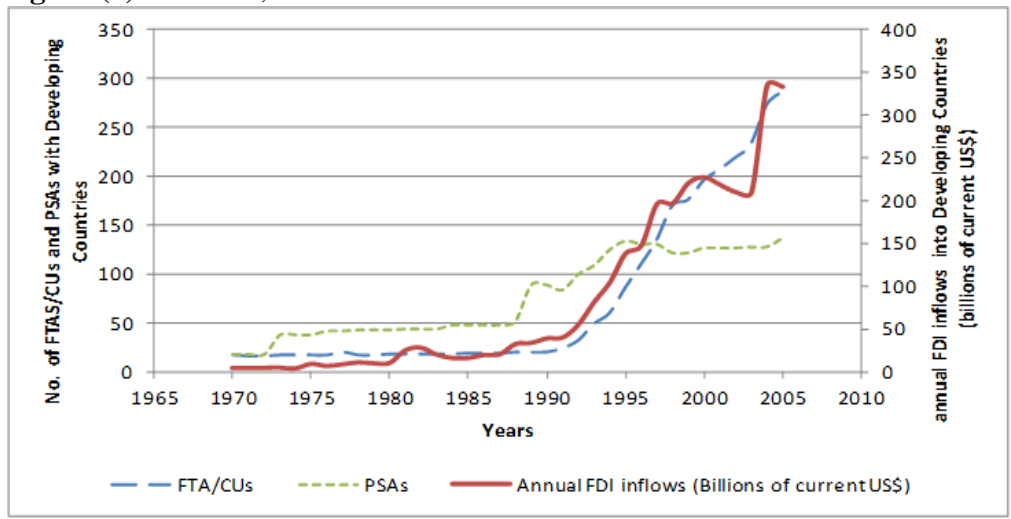

## *2.2. Heterogeneous International institutions, domestic politics and credibility of commitment in attracting FDI*

The impact of democratic regime type is positive in attracting FDI (Busse 2003). In the same vein, Jensen (2003) has argued that democratic political institutions at the domestic level constrain the committed policy flexibility of the executive. Compared to regimes tending towards autocracy, democratic regimes are characterized by heterogeneous veto players which weaken the ability of the executive to back on ex-ante commitments. Democracies signal their intention to other states and are able to commit credibly and clearly as opposed to authoritarian states (Fearon 1994, and Pevehouse *et al.* 2002). Li and Resnick (2003) have argued that property rights are stronger in democratic regimes, thus providing confidence to investors that they are safe from expropriation of their assets in host countries. However, this argument could also be contradicted in literature. For example, Li (2009) argued that democratic regime is not a remedy for eliminating the risk of expropriation. According to Yang (2007) there is no systematic relationship between democracy and FDI inflows. Being a democracy does not help to attract higher levels of FDI. Analyzing the arguments of Jensen (2003) and Li and Resnick (2003), he argued that these studies do not offer complete theoretical explanations on how political institutions influence expropriation of foreign assets i.e. they did not analyze opportunistic incentives on part of politicians.

The debate is going on and indecisive, whether democratic countries are credible in respecting the rights of foreign investors and attract FDI, there is another mechanism through which countries can assure investors and attract FDI is by concluding bilateral and regional trade agreements. As argued, trade agreements provide mainly two direct advantages in the context of investments: they include, in general, investment provisions and at the same time, provisions to settle dispute (if raised) related to investments.[7] Thus, through these institutionalized arrangements, the democratic governments ensure the credibility of their policy commitment (ex-post) to foreign investors (Büthe and Milner 2010). In this vein, it is important to understand the dynamics of regime type and domestic veto players in the formation of varying types of trade agreements with the goal to attract foreign firms (ex-ante) and the potential for reneging on agreements concluded. A problem arises at domestic level, when democratic governments pursue the policy of entering into trade agreements, especially the type of agreement proposed. Of crucial importance in this regard are the differences in formal institutions for sharing decision-making power that create the potential for veto players (Mansfield *et al.* 2008). The democratic political regime has to assume the conflicting interests of domestic institutions before concluding trade agreements. These players are the institutional and partisan actors whose assent is necessary to introduce new policies (Tsebelis 2002). In a coalition government, any one member may veto a certain policy proposal by threatening to withdraw from government if there are certain reservations. In general, the domestic actors are divided into two main categories; median voters[8] and interest groups. The results of Mansfield *et al.* (2008) indicate that more the domestic veto players, the higher the probability that the governments will form lower/shallower forms of integration agreements. But the preferences of domestic institutions are diverse i.e. there exists strong heterogeneity in domestic groups as well. Stolper Samuelson theorem predicts that free trade is more beneficial to the working class in labor-abundant countries. Here the workers are winners and capitalists are losers; and vice versa in capital-rich countries. In an unequal country, the median-voter will be labor-rich and capital poor thus it will influence government to adopt pro-labor policies. Median voter approach (pro-labor) is assumed by Levy (1997) in his widely studied political economy model based on Hecksher-Ohlin framework. Other competing theory in contrast to median voter theory, the lobbying model, also based on two-factor Hecksher-Ohlin framework, Rodrick (1986) argues that capitalist lobby is more concentrated (and possessing better information (Goldstein and Martin 2000)) than workers and therefore are more effective in pressurizing the government to conclude the policy which would be more pro-capital thus giving protection to the possessors of capital in labor-abundant countries[9].

---

[7] The indirect effect of trade agreements on FDI can be seen from the fact that these arrangements lower tariff barriers leading to increased trade flows. FDI could be closely related to trade flows, either as complementary such as intra-firm or inducing substitution to trade.

    The heterogeneity, described above, among domestic groups gives rise to the alternate choices of legalization and procedures in negotiations of arrangements. The import-competing groups feel themselves threatened as they face potential loss of market share in the presence of foreign firms and production capital. They restrain the government to commit the highly legalized form of arrangement thus falling short of full-fledged arrangement. On the other hand, in the context of developing countries, the state has to fulfill the demands of median voters, which encourage governments of committing highly legalized agreements, also in order to get them reelected as the median voters constitute a larger portion of population. Foreign direct investment is favorable for other part of median voters i.e. consumers in the host country as they help in improving the quality of products from the resulting increased competition thus benefitting consumers. FDI also helps in: technology advancement; getting access to wider global market by integrating to the world economy thus benefiting exporters[10]; improved human resources permitting labor force to acquire globally valued skills (promoting general interests and benefiting median voter). In the same vein, as argued that according to Stolper Samuelson theorem, free trade (under the auspices of deeper trade agreements FTA-CU) is beneficial for

---

[8] Median voters are sometimes called 'selectorates' mentioned by (Mansfield *et al.* 2008) as they make up a broader portion of society in a democracy and choose the leader and keep them in office.
[9] Labor-rich countries are developing countries, which are the focus of our analysis here, who obtain benefits from FDI, which comes through institutionalized arrangements providing more credibility.
[10] Investments are encouraged in trade agreements and for that matter the design of arrangement is important. The argument is relating trade and investment; the inclusion of wide range of products is entailed in the formation of deeper agreements.

labor abundant (developing) countries. In other words, the democratic government in the developing country will be more interested in concluding deeper forms of arrangements, which would be utile for welfare of their median voters (labor abundant audience) and also guaranteeing the international investors.

The credibility of commitment by a democratic government, described above, follows the sustainability of policies committed in concluded arrangement. More legalized trade regime will provide more and better information about the domestic distributional implications of trade agreements. This information will mobilize the domestic players in democratic regime to act in favor or against the arrangement (Goldstein and Martin 2000). They further argued that better information will empower protectionists when the governments are in process of concluding trade agreements and free traders on the issues of compliance to existing agreements. Reneging on commitments to FDI may be politically beneficial in the short run as increased income flows from policy changes can be distributed to a small group comprising core political supporters (Jensen 2003). Political leaders realize that liberalization (in the context of trade arrangements) will impose domestic costs and therefore they want to retain discretion to respond in future to uncertain demands for relief from injured groups (Smith 2000). Thus, the dynamics of domestic politics create some optimal level of imperfection in the application of international rules (Downs and Rocke 1995). The compliance with concluded agreements is enforced by other groups also such as median voters. As they constitute a large part of population, therefore cannot be ignored in democratic regimes.

Trade agreements increase domestic costs in the event of contract breach (Mitchell and Hensel 2007, and Tomz 2007) thus enhancing the reliability of the host governments that they will work for the continuity of policies. Also, the cost of reneging on commitment increases with legalization where the potential of opportunistic behavior is high (Abbott and Snidal 2000). Therefore, we argue that domestic cost of reneging on the deeper commitments such as FTA-CU will be higher than the PSA. Deeper trade agreements offer the governments to gain credibility (by increasing costs and locking-in) both at domestic and international level. There exist two types of costs *to be beared* by a democratic government in the event of noncompliance: International reputational costs (Mitchell and Hensel 2007, and Smith 2000) and Domestic audience cost (Fearon 1994, and Tomz 2007). Since the governments, where the regime type is democratic, seek the support of median voter because they find it more useful to comply with deeper and broader arrangements as the median voter is more concerned (with actions of government) here with respect to the partial arrangements. As anecdotal evidence for this argument, Tomz find the support based on the experimental surveys.

The domestic politics play a role in developed countries also whose investors make investments abroad also in ensuring credibility. The barriers to trade and investment restrict the prospects of benefitting from differences in capital costs between countries. Chase (2003) finds the support for the argument that reducing barriers through regional trading agreement significantly help investors to move their production facilities. Thus, the investors, being part of the domestic pressure groups and in order to benefit from larger market and at the same time to secure their investments, force their governments to conclude the deeper type of agreement ascertaining the high credibility of policy commitment. Koremenos *et al.* (2001) argued that non-state actors participate with increasing frequency in institutional design. They further argued that trade agreements are self conscious creation of states (and to a lesser extent, of interest groups and corporations). Arguing on the same logic but regarding the domestic institutions of U.S., Bhagwati (2008) maintains that ethnic groups and bureaucracies pushed for particular PTAs. The above arguments lead us to the following hypothesis:

H2: The more the country tends towards democracy, the significant positive effect of FTA-CU on FDI inflows rises whereas the effect of PSA remains insignificant.

### 3. Empirical Analysis

#### 3.1. Data

The sample comprises an unbalanced panel of 122 Non-OECD developing countries from the 1970 to 2005. The data for the dependant variable, FDI inflows as a percentage of GDP is taken from the online version of UNCTAD's Handbook of Statistics. The primary source of the data on Partial Scope Agreements and Free Trade Agreements/Custom Unions is WTO[11]. There exist many agreements that have not been notified to WTO. For that matter, the comprehensive database of preferential trade agreements compiled at Mc Gill University is consulted[12]. Data on variable POLITY which signifies countries' tendency towards democracy is taken from Polity IV Project. Number of Bilateral Investment Treaties signed by specific country is taken from UNCTAD[13]. Data on Political Constraints is extracted from the database compiled by Henisz Witold. Data on Population and GDP per capita is accessed through Penn World Table. GDP growth variable is calculated from GDP variable in constant (2005) dollars. GDP per capita is also in constant (2005) dollars. The descriptive statistics for variables can be seen in Table (1).

**Table (1)** DESCRIPTIVE STATISTICS

| Variables | Mean | Std. Deviation | Min | Max |
|---|---|---|---|---|
| FDI | 2.095 | 4.798 | -65.411 | 92.104 |
| Cumulative FTA-CU | 0.609 | 1.478 | 0 | 17 |
| Cumulative PSA | 0.631 | 0.967 | 0 | 7 |
| Democracy (Polity) | -1.056 | 6.828 | -10 | 10 |
| GATT-WTO Membership | 0.547 | 0.498 | 0 | 1 |
| Cumulative BIT | 7.279 | 13.456 | 0 | 107 |
| Log (Population) | 15.921 | 1.411 | 12.292 | 20.990 |
| GDP Growth | 3.960 | 8.563 | -66.972 | 168.008 |
| GDP per capita | 8.101 | 0.984 | 5.032 | 11.491 |

**Notes**: (FDI): Foreign Direct Investment, (FTA-CU): Free Trade Agreements and Customs Unions, (PSA): Partial Scope Agreements, (Polity): democratic level, (GATT-WTO): membership in international organizations (General Agreement on Tariffs and Trade - World Trade Organization), (BIT): Bilateral Investment Treaties. All variables are de-meaned by country and de-trended. (See section 4 for details). These statistics are based on one-year lagged values.

#### 3.2. Methodology and Results

Following the introduction of variables, we turn to the empirical linkages between the regime type, trade agreements and FDI flows. The fixed-effect panel analysis comprising 122 developing countries and the entire period of 1970-2005 is performed. The rationale for using fixed effects is that they control for omitted unobservable factors which is source of endogeneity bias. There are two distinct instances of omitted unobservable variables. First, the countries may self select into trade agreements. As Downs, Rocke and Barsoom (1996) argue that governments tend to conclude only those agreements that oblige them to do what they are already doing or want to do. Second, there exist country specific characteristics which are constant over time. If these country characteristics have an impact on the FDI inflows and one or more explanatory variables (notably the trade agreements) then the structural disturbance (which captures heterogeneity across units of observation) will be correlated with those explanatory variables. Unlike other estimation techniques, for example OLS on cross-section data that produce biased estimates or random-effects models that assume no correlation between unobservables ($a_{i,t}$) and the variables of cumulative FTA-CU and cumulative PSA variables, the fixed effects estimator is an unbiased and consistent estimator of the treatment effect of trade agreements.

---

[11] http://www.wto.org/english/tratop_e/region_e/region_e.htm
[12] http://ptas.mcgill.ca/
[13] http://www.unctad.org/Templates/Page.asp?intItemID=2344&lang=1

In order to estimate the treatment effect of heterogeneous trade agreements on FDI flows into developing countries and the conditional effect of regime type on these heterogeneous agreements, the following equation is estimated:

$$FDI_{i,t} = \beta_0 + \beta_1 FTA\_CU_{i,t-1} + \beta_2 PSA_{i,t-1} + \beta_3 POLITY_{i,t-1} + \beta_4 BIT_{i,t-1} + \beta_5 GATT\_WTO_{i,t-1} + \beta_6 FTA\_CU * POLITY_{i,t-1} + \beta_7 PSA * POLITY_{i,t-1} + \beta_8 Control_{i,t-1} + a_{i,t} + \varepsilon_{i,t-1}$$

The dependant variable is in fact the FDI as percentage of GDP. As can be seen all the variables are lagged. The variables showed the trend; therefore they are de-trended also to avoid spurious correlation among the dependant and independent variables.

The results are presented in Table (2). We start the analysis from the impact of regime type in attracting FDI inflows in model (1) which is insignificant. Among the control variables of population, GDP Growth and GDP per capita, the variable GDP Growth is positive and significant at 10 percent level. In estimation (2), the effect of international institutions is added in model (1). In the presence of control variables and regime type, the effect of Bilateral Investment Treaties and GATT-WTO membership is significantly positive at 10 percent level. More the country concludes BIT, the more the FDI inflows it experiences. GATT-WTO membership also increases the prospects for a country to attract investments. In model (3), two different types of trade agreements are included such as Free Trade Agreements-Custom Unions (FTA-CU) and Partial Scope Agreements (PSA). The impact of FTA-CU is positive and statistically significant at 10 percent level whereas PSA are also showing positive sign but insignificant. This confirms our first hypothesis (H1) that by concluding more FTA-CU the country gains the confidence of foreign investors, signals its credibility of commitments and thus attract higher FDI inflows. The PSA do not play any role in attracting FDI. The variable of regime type (i.e. POLITY) remains insignificant showing that democracy itself does not signal to foreign investors about the security of their investments. The control variables are showing the same behavior. Population of a country and GDP per capita are insignificant though positive. The last model (4) shows the interaction effect of both types of trade agreements in addition to all other explanatory variables. The sign of the interaction variable between cumulative FTA-CU and POLITY is positive and significant at 5 percent level whereas the interaction effect of cumulative PSA and POLITY is not significantly different from zero. The result confirms our second hypothesis (H2). This shows that more the regime type tends towards democracy, the effect of full-fledged agreements such as FTA-CU is larger and significant to attract FDI. The democratic regime moderates significantly the effect of full-fledged trade agreements on FDI inflows whereas it does not condition the impact of partial scope agreements. The interaction effects can be shown through a graph. Figure 2 shows the effect of FTA-CU on FDI inflows moderated by the democratic level (Polity).

FDI inflows are boosted when a highly democratic country concludes more FTA-CU. Although the effect of low democracy level with larger amount of FTA-CU concluded is at the increasing trend but this is very small. Countries with the higher level of democracy and lower FTA-CU experience lower FDI inflows depicting the importance of FTA-CU in boosting FDI.

The marginal effect of FTA-CU and PSA are calculated for each of the variables conditional on one another along with 99 percent confidence intervals based on variance-covariance matrix (Brambor, Clark and Golder 2006). Figure (2) displays the marginal effect for FTA-CU, conditioned by regime type (polity) estimated in model (4). It clearly shows that FTA-CU do not have significant effect on FDI in the countries where polity score is low (i.e. countries having autocratic regime type), whereas, as the level of democracy rises, the FTA-CU show significant and positive effect on FDI. In figure (3), we analyzed the effect of domestic regime type conditioned by FTA-CU. Here, the effect of regime type on FDI is insignificant at any number of FTA-CU signed. In other words, the FTA-CU to which the country is party, does not condition the effect of regime type on FDI and the effect cannot be distinguished from zero at any point. Figure (4) shows the marginal effect of PSA on FDI moderated by regime type in country whereas figure (5) depicts the effect of polity on FDI moderated by PSA. Both figures prove that the effects are insignificant. Partial scope agreements do not have significant effects on FDI moderated by polity.

**Table (2)** CONDITIONAL EFFECT OF REGIME TYPE AND HETEROGENEOUS TRADE AGREEMENTS

|  | (1) | | (2) | | (3) | | (4) | |
|---|---|---|---|---|---|---|---|---|
| Cumulative FTA-CU |  |  |  |  | **0.158*** | (0.088) | **0.285**** | (0.134) |
| Cumulative PSA |  |  |  |  | 0.140 | (0.188) | 0.299 | (0.185) |
| POLITY | 0.013 | (0.020) | 0.004 | (0.019) | 0.009 | (0.019) | 0.011 | (0.019) |
| Cumulative BIT |  |  | **0.022*** | (0.013) | 0.016 | (0.013) | 0.008 | (0.014) |
| GATT-WTO Membership |  |  | **1.181*** | (0.675) | 1.080 | (0.703) | 1.009 | (0.666) |
| Cumulative FTA-CU (i+1) |  |  |  |  |  |  | -0.069 | (0.107) |
| Population | -0.010 | (1.244) | 0.629 | (1.118) | 1.083 | (1.265) | 0.522 | (1.328) |
| GDP Growth | **0.055*** | (0.032) | **0.055*** | (0.032) | **0.054*** | (0.032) | 0.056 | (0.034) |
| GDP per capita | 0.577 | (0.651) | 0.222 | (0.682) | 0.235 | (0.700) | 0.275 | (0.748) |
| Cumulative FTA-CU * POLITY |  |  |  |  |  |  | **0.053**** | (0.024) |
| Cumulative PSA * POLITY |  |  |  |  |  |  | -0.009 | (0.020) |
| Constant | -0.000 | (0.027) | -0.000 | (0.028) | -0.000 | (0.029) | -0.038 | (0.043) |
| No. of Countries | 122 | | 122 | | 122 | | 119 | |
| Observations | 3565 | | 3565 | | 3565 | | 3330 | |
| R-squared | 0.015 | | 0.022 | | 0.024 | | 0.027 | |

**Notes**: (FTA-CU): Free Trade Agreements and Customs Unions, (PSA): Partial Scope Agreements, (Polity): democratic level, (GATT-WTO): membership in international organizations (General Agreement on Tariffs and Trade - World Trade Organization), (BIT): Bilateral Investment Treaties. Regression with robust standard errors for correction of heteroscedacity in parantheses. ** denotes significance at 5 percent; * significant at 10 percent. All variables are de-meaned, de-trended and one-year lagged.

The strict exogeneity of the results is also tested by using a method suggested by Baier and Bergstrand (2007). For this, future level of cumulative FTA-CU is introduced in the regression model. If the changes of this independent variable are strictly exogenous to FDI flow changes, cumulative FTA-CU$_{t+1}$ should be uncorrelated with the FDI flow and so the coefficient must be insignificant. The result of this variable confirms this and therefore there is no feedback effect.

**Figure (2)** MARGINAL EFFECT OF FTA-CU AS A FUNCTION OF POLITY

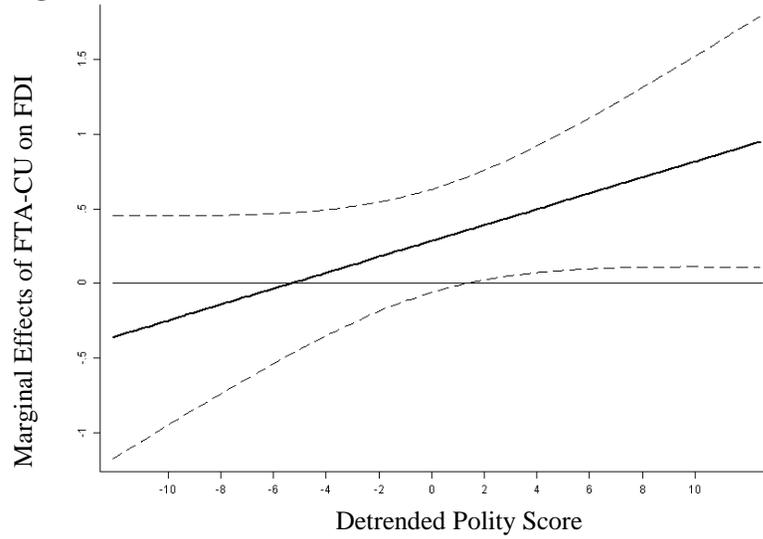

**Figure (3)** MARGINAL EFFECT OF POLITY AS A FUNCTION OF FTA-CU

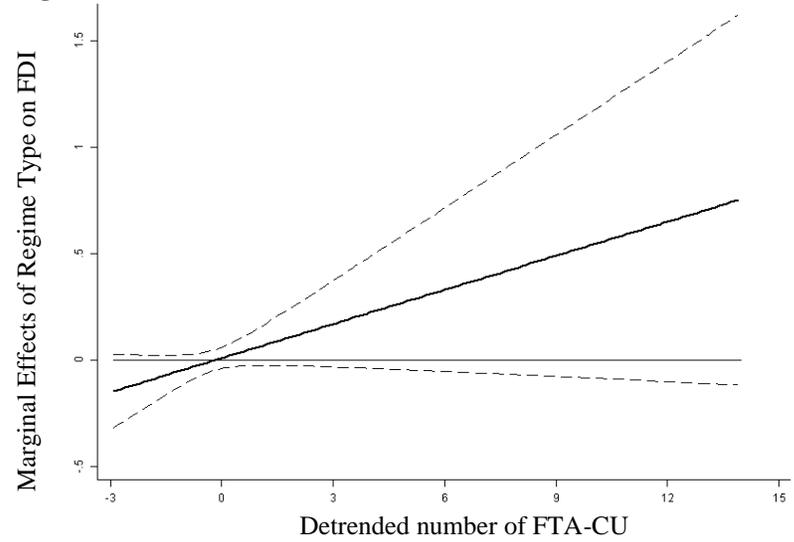

**Figure (4)** MARGINAL EFFECTS OF PSA AS A FUNCTION OF POLITY

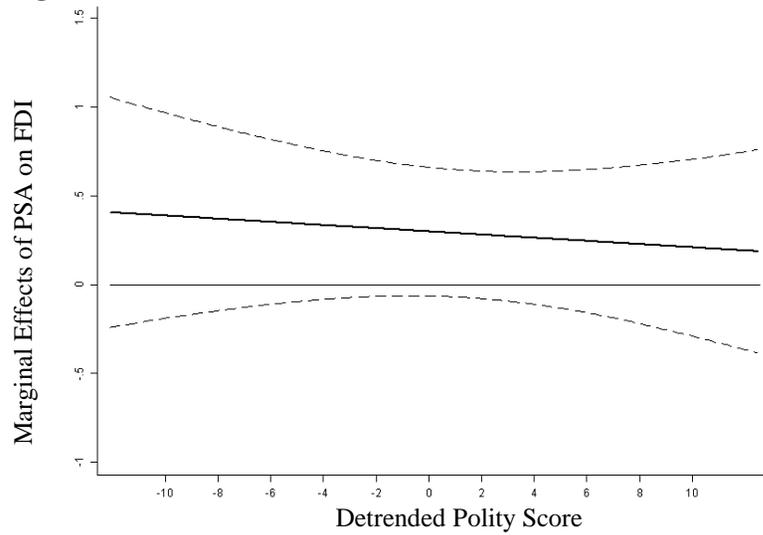

**Figure (5)** MARGINAL EFFECT OF POLITY AS A FUNCTION OF PSA

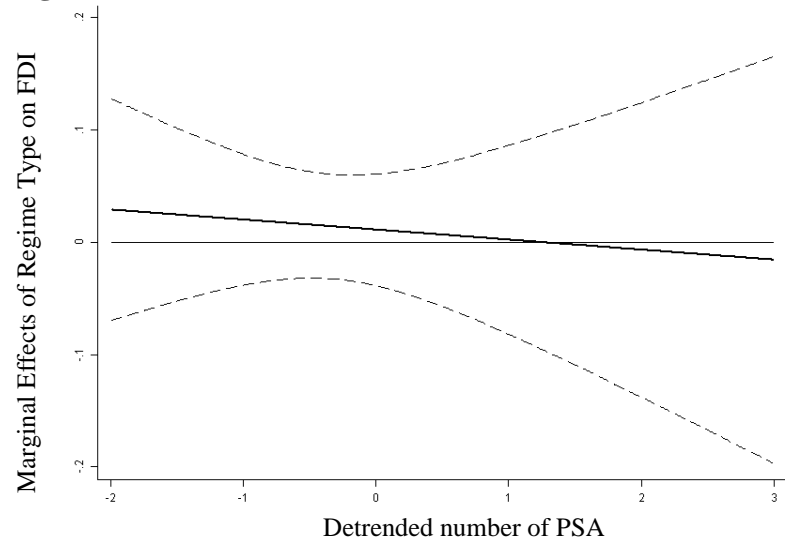

## 4. Concluding Remarks

Foreign Direct Investment (FDI) inflows have recently been the most persistent source of capital inflows in developing countries. This extensive growth of FDI has, consequently, given rise to the competition among policy makers in developing countries and they pursue to conclude international trade agreements to boost foreign direct investment inflows. But, the trade agreements are heterogeneous in nature and there is difference on their impact in attracting FDI. Full-fledged agreements signal credibility of commitments. Also there is debate whether the level of democracy encourages foreign investors to invest. The goal of this paper was to explore in detail the role of heterogeneous international trade agreements in their role to attract FDI flows and to examine the conditional effect of level of democracy in the country on different types of agreements in order to attract FDI flows.

Accordingly, the results can be summarized as follows: In the fixed-effects panel analysis, covering a period of 36 years, the impact of full-fledged agreements is found to be significant and positive. They signal the credibility of the host government. At the same time, the impact of partial scope agreements is proved to be insignificant. The level of democracy itself is insignificant but its conditional effect on full-fledged agreements is significant. The moderating effect of democracy on partial scope agreements is found to be insignificant. This shows that democratic regimes do care more about their commitments when they conclude the agreements ascertaining strong dispute settlement mechanisms. These agreements increase the reputational cost for democratic regimes in case of reneging from their commitments. The impact of other international institutions (Bilateral Investment Treaties and GATT-WTO membership) is found to be positive and significant in the earlier estimations but after the introduction of trade agreements, these institutions lost their significance which could be a signal that countries rely more on trade agreements to boost investment rather than other types of institutions. The strict exogeneity of the results is also tested by using a method suggested by Baier and Bergstrand (2007) and no feedback effect is found.

This research could be continued by introducing individual clauses of trade agreements regarding investment and the level of legalization as democracies are more prone to respect higher commitments. From that one could investigate the heterogeneity in trade agreements and analyze their effect on FDI.

**References**

Abbott, F.M. (2000) "NAFTA and the Legalization of World Politics: A Case Study" *International Organization* **54**, 519-547.

Abbott, K.W and D. Snidal (2000) "Hard and Soft Law in International Governance" *International Organization* **54**, 421-456.

Abbott, K.W and D. Snidal (1998) "Why states act through Formal International Organizations" *The Journal of Conflict Resolution* **42**, 3-32.

Baier, S.L and J.H. Bergstrand (2007) "Do free trade agreements actually increase members' international trade?" *Journal of International Economics* 7**1**, 72-95.

Balassa, B. (1961) "Towards a Theory of Economic Integration" *Kyklos* **16**, 1-17.

Bhagwati, J. (2008) *Termites in the Trading System: How Preferential Agreements Undermine Free Trade*, Oxford: Oxford University Press.

Blomström, M and A. Kokko (1997) "Regional integration and foreign direct investment" NBER working paper number 6019.

Brambor, T. Clark, W.R. and M. Golder (2006) "Understanding Interaction Models: Improving Empirical Analyses" *Political Analysis* **14**, 63-82.

Busse, M. (2003) "Democracy and FDI" HWWA discussion paper no. 216, HWWA, Hamburg.

Büthe, T and H.V. Milner (2008) "The Politics of Foreign Direct Investment into Developing Countries: Increasing FDI through International Trade Agreements?" *American Journal of Political Science* **52**, 741-762.

Büthe, T and H.V. Milner (2010) "The Interaction of Domestic and International Institutions: Democracy, Preferential Trade Agreements, and Foreign Direct Investment into Developing Countries" Paper presented at the Workshop on the Politics of Preferential Trade Agreements: Theory, Measurement, and Empirical Applications, Princeton University, Niehaus Center for Globalization and Governance May 1.

Chase, K.A. (2003) "Economic Interests and Regional Trading Arrangements: The Case of NAFTA" *International Organization* **57**, 137-1740.

Downs, G.W and D.M. Rocke (1995) *Optimal Imperfection? Domestic Uncertainty and Institutions in International Relations*, Princeton, N.J.: Princeton University Press.

Downs, G.W. Rocke D.M. and P.N. Barsoom (1996) "Is the good news about compliance good news about cooperation?" *International Organization* **50**, 379-406.

Fearon, J. (1994) "Domestic political audiences and the escalation of international disputes" *American Political Science Review* **88**, 577-92.

Fernandez, R and J. Portes (1998) "Returns to Regionalism: An Analysis of Nontraditional Gains from Regional Trade Agreements" *World Bank Economic Review* **12**, 197-220.

Goldstein, J and L.L. Martin (2000) "Legalization, Trade Liberalization, and Domestic Politics: A Cautionary Note" *International Organization* **54**, 603-632.

Henisz, W.J. (2000) "The Institutional Environment for Multinational Investment" *Journal of Law, Economics and Organization* **16**, 334-364.

Jensen, N.M. (2003) "Democratic Governance and Multinational Corporations: Political Regimes and Inflows of Foreign Direct Investment" *International Organization* **57**, 587-616.

Kahler, M. (2000) "Legalization as Strategy: The Asia-Pacific Case" *International Organization* **54**, 549-571.

Koremenos, B. Lipson, C and D. Snidal (2001) "The Rational Design of International Institutions" *International Organization* **55**, 761-799.

Levy, P. (1997) "A political-economic analysis of free-trade agreements" *American Economic Review* **87**, 506-519.

Li, Q. (2009) "Democracy, autocracy, and expropriation of foreign direct investment" *Comparative Political Studies* **42**, 1098-1127.

Li, Q and A. Resnick (2003) "Reversal of fortunes: Democracy, property rights and foreign direct investment inflows in developing countries" *International Organization* **57**, 1-37.

Mansfield, Edward D., Helen V. Milner and Jon C. Pevehouse. 2008. Democracy, Veto Players and the Depth of Regional Integration. *World Economy* 31, 1: 67-96.

Mitchell, S.M. and P.R. Hensel (2007) "International Institutions and Compliance with Agreements" *American Journal of Political Science* **51**, 721-737.

Pevehouse, J.C. Hafner-Burton E and M. Zierler (2002) "Regional Trade and Institutional Design: Long After Hegemony?" Presented at the 2002 MPSA Meetings.

Schiff, M and L.A. Winters (1998) "Regional Integration as Diplomacy" *The World Bank Economic Review* **12**, 271-295.

Simmons, B.A. (2000) "The Legalization of International Monetary Affairs" *International Organization* **54**, 573-602.

Smith, J.M. (2000) "The Politics of Dispute Settlement Design: Explaining Legalism in Regional Trade Pacts" *International Organization* **54**, 137-180.

Stasavage, D. (2002b) "Private Investment and Political Uncertainty" *Economics and Politics* **14**, 41-63.

Tomz, M. (2007) "Domestic Audience Costs in International Relations: An Experimental Approach" *International Organization* **61**, 821-840.

Tsebelis, G. (2002) *Veto Players: How Political Institutions Work*, Princeton: Princeton University Press.

UNCTAD (2011) *World Investment Report 2011*, New York: United Nations Press.

Yang, B. (2007) "Autocracy, Democracy, and FDI Inflows to the Developing Countries" *International Economic Journal* **21**, 419-439.